\documentclass[lettersize,journal]{IEEEtran}
\usepackage{amsmath,amsfonts}
\usepackage{algorithmic}
\usepackage{algorithm}
\usepackage{array}
\usepackage[caption=false,font=normalsize,labelfont=sf,textfont=sf]{subfig}
\usepackage{textcomp}
\usepackage{stfloats}
\usepackage{url}
\usepackage{bm}
\usepackage{verbatim}
\usepackage{multirow}
\usepackage{graphicx}
\usepackage{cite}
\usepackage{threeparttable}
\usepackage[normalem]{ulem}
\usepackage{newtxtext}
\usepackage[switch]{lineno}

 \usepackage{booktabs}
 \usepackage{amssymb}
 \usepackage{bbm}  
 \usepackage{enumitem}
\usepackage{placeins}

\newcommand{\dataset}{\textsc{MDUR}}

\begin{document}

\title{Compositional-Degradation UAV Image Restoration: Conditional Decoupled MoE Network and A Benchmark}

\author{Jinquan Yan, Zhicheng Zhao*, Zhengzheng Tu, Chenglong Li, Jin Tang, and Bin Luo
\thanks{* Corresponding author: Zhicheng Zhao.}
\thanks{This work was supported in part by the National Natural Science Foundation of China (No. 62306005, 62006002, 62076003, and 62576006), and in part by the Natural Science Foundation of Anhui Higher Education Institution (No. 2022AH040014).}

\thanks{Zhicheng Zhao, and Chenglong Li are with Key Laboratory of Intelligent Computing \& Signal Processing (Anhui University), Ministry of Education, Anhui Provincial Key Laboratory of Multimodal Cognitive Computation, School of Artificial Intelligence, Anhui University, Hefei 230601, China. (Email: zhaozhicheng@ahu.edu.cn, lcl1314@foxmail.com).}

\thanks{Jinquan Yan, Zhengzheng Tu, Jin Tang and Bin Luo are with Anhui Provincial Key Laboratory of Multimodal Cognitive Computation, School of Computer Science and Technology, Anhui University, Hefei 230601, China. (Email: jinquanyan001@gmail.com,07036@ahu.edu.cn, tangjin@ahu.edu.cn, luobin@ahu.edu.cn).}
}

\markboth{Yan \MakeLowercase{\textit{et al.}}: Compositional-Degradation UAV Image Restoration: Conditional Decoupled MoE Network and A Benchmark}%
{Yan \MakeLowercase{\textit{et al.}}: Compositional-Degradation UAV Image Restoration: Conditional Decoupled MoE Network and A Benchmark}

\maketitle
\begin{abstract}
UAV images are critical for applications such as large-area mapping, infrastructure inspection, and emergency response. However, in real-world flight environments, a single image is often affected by multiple degradation factors, including rain, haze, and noise, undermining downstream task performance. 
Current unified restoration approaches typically rely on implicit degradation representations that entangle multiple factors into a single condition, causing mutual interference among heterogeneous corrections. 
To this end, we propose DAME-Net, a Degradation-Aware Mixture-of-Experts Network that decouples explicit degradation perception from degradation-conditioned reconstruction for compositional UAV image restoration. 
Specifically, we design a Factor-wise Degradation Perception module(FDPM) to provide explicit per-factor degradation cues for the restoration stage through multi-label prediction with label-similarity-guided soft alignment, replacing implicit entangled conditions with interpretable and generalizable degradation descriptions. 
Moreover, we develop a Conditioned Decoupled MoE module(CDMM) that leverages these cues for stage-wise conditioning, spatial-frequency hybrid processing, and mask-constrained decoupled expert routing, enabling selective factor-specific correction while suppressing irrelevant interference. 
In addition, we construct the Multi-Degradation UAV Restoration benchmark (MDUR), the first large-scale UAV benchmark for compositional UAV image restoration, with 43 degradation configurations from single degradations to four-factor composites and standardized seen/unseen splits.
Extensive experiments on MDUR demonstrate consistent improvements over representative unified restoration methods, with greater gains on unseen and higher-order composite degradations. Downstream experiments further validate benefits for UAV object detection. The dataset is available at \url{https://github.com/mmic-lcl/Datasets-and-benchmark-code}.
\end{abstract}

\begin{IEEEkeywords}
Image Restoration, Compositional Restoration, Mixture-of-Experts
\end{IEEEkeywords}

\begin{figure}
    \centering
    \includegraphics[width=\linewidth]{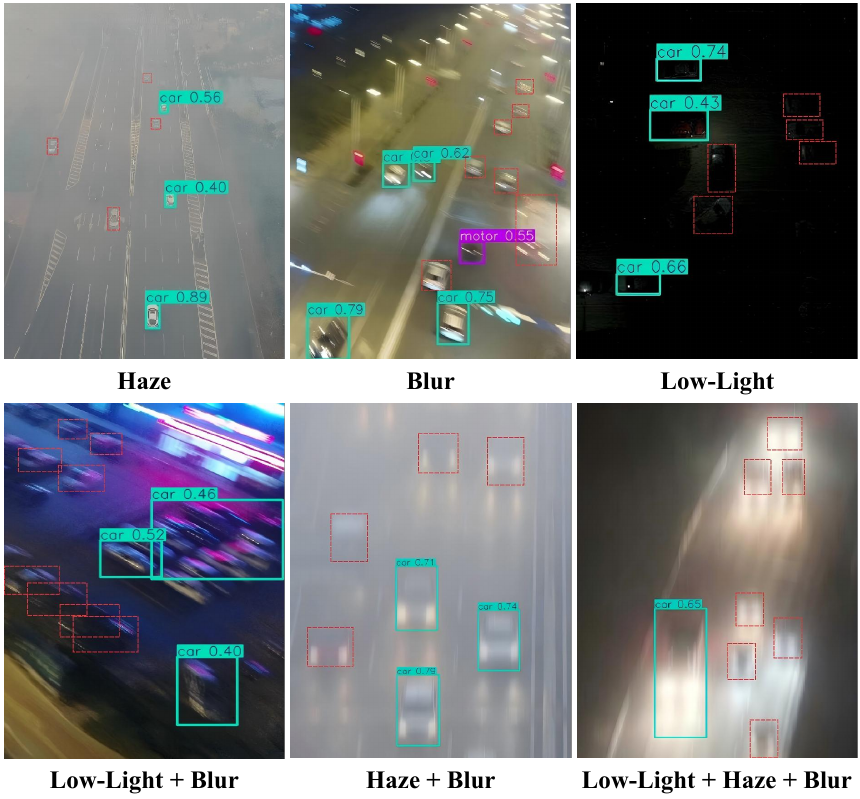}
    \caption{Real UAV examples illustrating the impact of adverse degradations on downstream object detection. The examples include single-degradation cases and composite cases. Colored solid boxes denote detector predictions, and red dashed boxes highlight visually identifiable missed targets. Adverse degradations substantially reduce detection reliability, with more severe failures in composite cases.}
    \label{wrong_detect}
\end{figure}

\section{Introduction}
\label{sec:intro}

UAV image restoration matters in applications such as large-area mapping, infrastructure inspection, environmental monitoring, and emergency response, where both human analysis and downstream perception depend on image quality~\cite{li2022ultrahigh, pan2022dual}. In real flight conditions, however, UAV images are rarely affected by only one corruption. Haze, rain, snow, low-light, over-exposure, motion blur, sensor noise, and compression artifacts often co-occur in a single image~\cite{munir2024impact, si2025dc}. These composite degradations are not just a sum of single-factor cases, as their interactions jointly distort structure, contrast, and local texture, making restoration harder~\cite{mao2024allrestorer}.


This difficulty also propagates to downstream perception. As shown in Fig.~\ref{wrong_detect}, real UAV examples under both single and composite degradations indicate that adverse conditions can severely impair a pre-trained YOLOv8n detector~\cite{varghese2024yolov8}, leading primarily to missed detections and reduced detection reliability, with more severe failures in composite cases~\cite{wang2025dual}. Compositional UAV restoration should therefore be treated as a dedicated problem rather than as a straightforward extension of single-degradation recovery.

\begin{figure}[t]
    \centering
    \includegraphics[width=\linewidth]{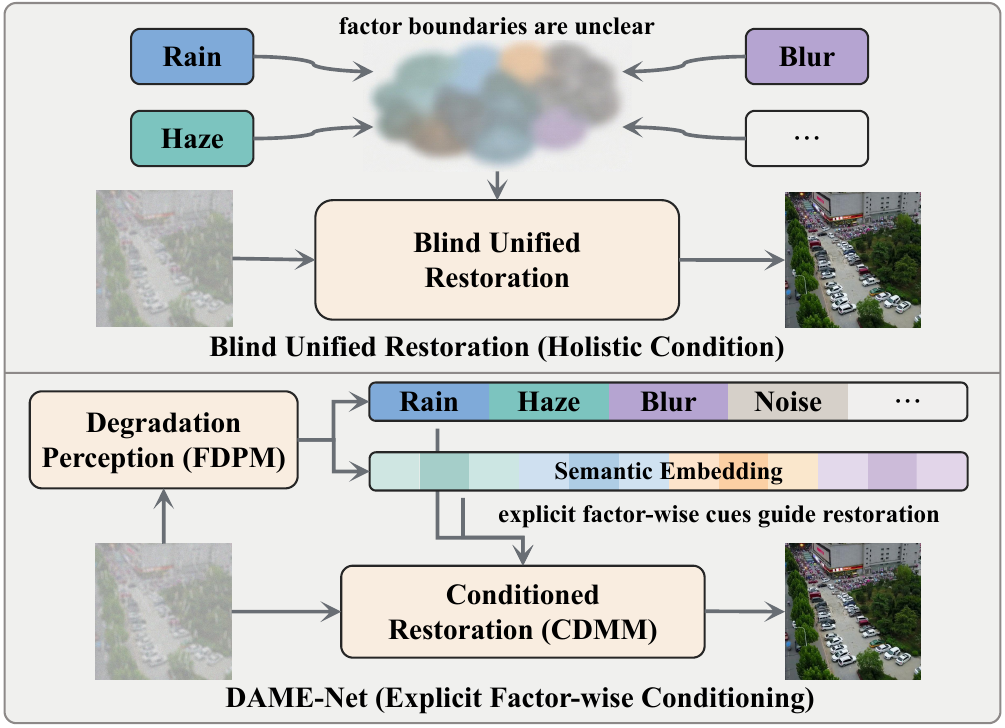}
    \caption{Motivation of DAME-Net. In blind unified restoration, multiple active degradations are often compressed into a single holistic condition, which blurs factor boundaries and limits factor-specific correction. DAME-Net instead performs explicit degradation perception and uses factor-wise cues for conditioned restoration.}
    \label{blind_restoration}
\end{figure}

Existing restoration methods do not directly solve this setting. Task-specific models for dehazing, deraining, low-light enhancement, and denoising~\cite{gui2023comprehensive, tian2023survey} work well when the degradation type is known, but their assumptions break down once several factors appear together~\cite{dong2026phydae}. Unified blind restoration methods avoid explicit degradation labels by learning a direct mapping from degraded to clean images~\cite{lin2024diffbir}. In the compositional setting, however, their implicit degradation representations often compress multiple active factors into a single holistic condition, making the conditioning signal too coarse to distinguish the underlying factor composition. As illustrated in Fig.~\ref{blind_restoration}, such entangled representations blur factor boundaries and hinder selective factor-specific correction. Although Mixture-of-Experts (MoE) methods partially introduce parameter separation~\cite{ren2024moe, lin2024unirestorer}, their routing is still typically implicit and may lack stable correspondence with specific degradation factors.

This limitation becomes more evident when composite degradations share partial factor overlap; for example, rain+haze is semantically related to both pure rain and pure haze rather than being an entirely independent category~\cite{wang2023promptrestorer}. These observations suggest that compositional restoration should be conditioned on explicit per-factor descriptions instead of an opaque holistic label.

\begin{figure}[t]
    \centering
    \includegraphics[width=\linewidth]{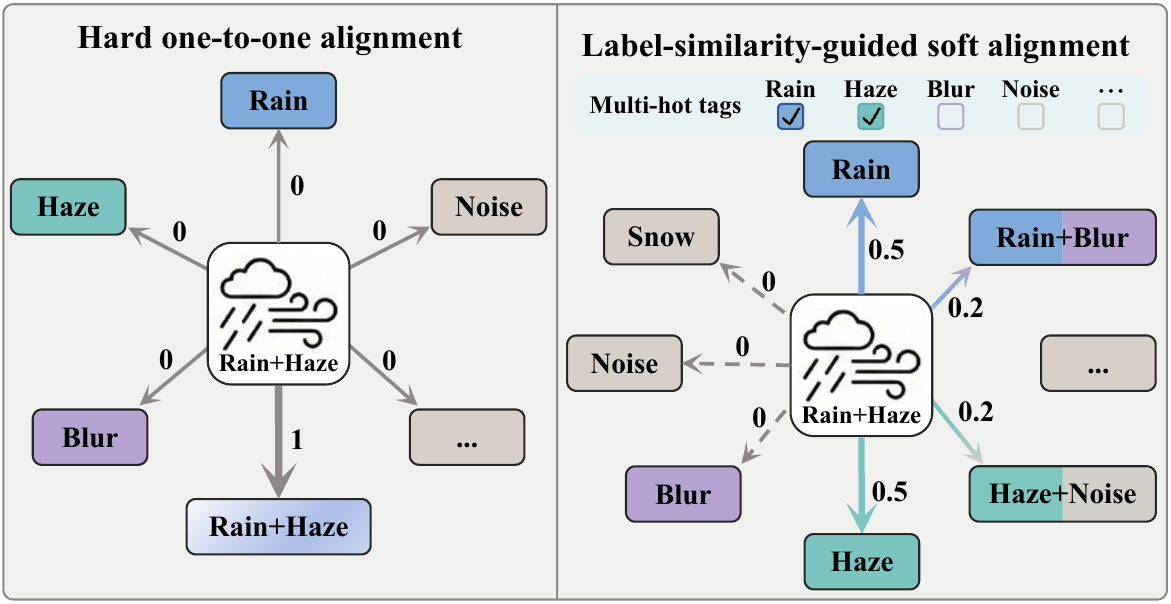}
    \caption{Motivation for FDPM. Conventional hard alignment forces a composite degradation (e.g., rain+haze) to match a single prompt, whereas FDPM preserves its overlap with constituent factors through label-similarity-guided soft alignment.}
    \label{sgdp_semantic}
\end{figure}

Motivated by these challenges, we propose DAME-Net, a Degradation-Aware Mixture-of-Experts Network that decouples explicit degradation perception from degradation-conditioned reconstruction for compositional UAV image restoration. Specifically, we design a Factor-wise Degradation Perception module (FDPM), a CLIP-based~\cite{radford2021learning} multi-label detector that provides explicit per-factor degradation cues through multi-label prediction with label-similarity-guided soft alignment, replacing implicit entangled conditions with interpretable and generalizable degradation descriptions, as illustrated in Fig.~\ref{sgdp_semantic}. Moreover, we develop a Conditioned Decoupled MoE module (CDMM) that leverages these cues for stage-wise conditioning, spatial--frequency hybrid processing, and mask-constrained decoupled expert routing, enabling selective factor-specific correction while suppressing irrelevant interference. In addition, we construct \dataset{}, a large-scale UAV benchmark for compositional image restoration, with 43 degradation configurations from single degradations to four-factor composites and standardized seen/unseen splits.

Our contributions are:
\begin{itemize}
\item We propose DAME-Net, a Degradation-Aware Mixture-of-Experts Network for compositional UAV image restoration. FDPM provides explicit per-factor degradation cues, and CDMM performs degradation-conditioned restoration with mask-constrained decoupled expert routing for selective factor-specific correction.
\item We introduce \dataset{}, the first large-scale UAV benchmark for compositional image restoration, with 43 degradation configurations from single degradations to four-factor composites and standardized seen/unseen splits for evaluating compositional generalization.
\item Extensive experiments on \dataset{} demonstrate consistent improvements over representative unified restoration baselines, with larger gains on unseen and higher-order composite degradations. Downstream experiments further validate benefits for downstream UAV object detection.
\end{itemize}

The main paper focuses on the method, averaged results, qualitative comparisons, downstream detection, and ablations. \dataset{} construction, task-wise tables, detector evaluation, and complexity analysis are deferred to the supplementary material.

\section{Related Work}
\label{sec:related}

\subsection{Single-Degradation Restoration}
Single-degradation restoration methods assume that one dominant corruption type is known or can be handled in isolation, such as haze, rain/snow, low-light, blur, noise, or compression artifacts~\cite{zhai2023comprehensive}. Earlier learning-based approaches typically trained a dedicated model for each task~\cite{yan2024image}, while more recent CNN- and Transformer-based methods improve performance through stronger feature extraction, multi-scale interaction, and attention mechanisms~\cite{park2025lightweight, zamir2022restormer, song2023vision}. These methods have established strong baselines for their respective tasks and remain important building blocks for image restoration.

Most single-degradation methods are developed under the assumption that one dominant corruption is present in each input. This assumption is often violated in real UAV imagery, where multiple degradation factors may co-occur and jointly affect restoration quality~\cite{yu2024multi}. As a result, methods specialized for isolated degradations are difficult to transfer to compositional UAV restoration, which motivates us to explicitly model multi-factor degradation settings.

\subsection{Unified Restoration}
Unified restoration methods aim to handle multiple degradation types with a single model instead of maintaining task-specific parameters~\cite{potlapalli2023promptir, cui2025adair}. Existing designs improve task awareness in different ways, including shared-backbone multi-task learning~\cite{li2022all, valanarasu2022transweather}, degradation embeddings or learned prompts~\cite{ren2024moe, potlapalli2023promptir}, expert routing~\cite{lin2024unirestorer, zamfir2025complexity, wang2025m2restore, kim2020restoring, feijoo2025towards}, and frequency-aware or degradation-aware modulation~\cite{cui2025adair, wu2025freprompter, cui2023selective, tian2025degradation, gu2024mixed}. Representative conditioning signals include noise-level maps, blur kernels, handcrafted degradation parameters, and learned latent codes, which are injected through affine modulation, dynamic filters, cross-attention, or prompt-based adaptation~\cite{zhang2018ffdnet, zhang2020deep}. These methods show that one network can cover diverse restoration tasks and provide strong unified restoration baselines.

Unified restoration methods improve model versatility by handling multiple degradations within a shared framework, but this shared conditioning becomes increasingly coarse in compositional settings. When several degradation factors are active simultaneously, a single latent code, prompt, or routing signal may not describe them with sufficient specificity. This motivates our use of explicit factor-wise degradation cues to guide restoration under compositional UAV degradations.

\subsection{Compositional Degradation Restoration}
Compositional degradation restoration addresses cases where multiple degradation factors co-occur in the same image and must be handled jointly~\cite{mao2024allrestorer}. Recent studies have explored mixed or compositional restoration through degradation estimation, prompt-based conditioning, and adaptive modulation~\cite{guo2024parameter, li2024promptcir, wang2023promptrestorer, gu2024mixed}. Some methods also attempt to infer degradation information from the input before restoration~\cite{zhang2023all, liu2025dpmambair, hu2025universal}. In this context, vision--language models such as CLIP provide a useful prior because their shared embedding space can encode semantic relations among different degradation descriptions~\cite{radford2021learning, shao2025controlling, yang2025vision}.

Existing compositional restoration methods have begun to address multi-factor degradations, yet many still rely on compressed holistic conditions that do not fully reflect the underlying factor composition. This mismatch becomes more evident when multiple degradation types co-occur and should be represented in a structured manner. Meanwhile, benchmarks for compositional UAV restoration remain limited, which motivates us to develop both DAME-Net and the MDUR benchmark in a unified framework.

\begin{figure*}[t]
    \centering
    \includegraphics[width=\linewidth]{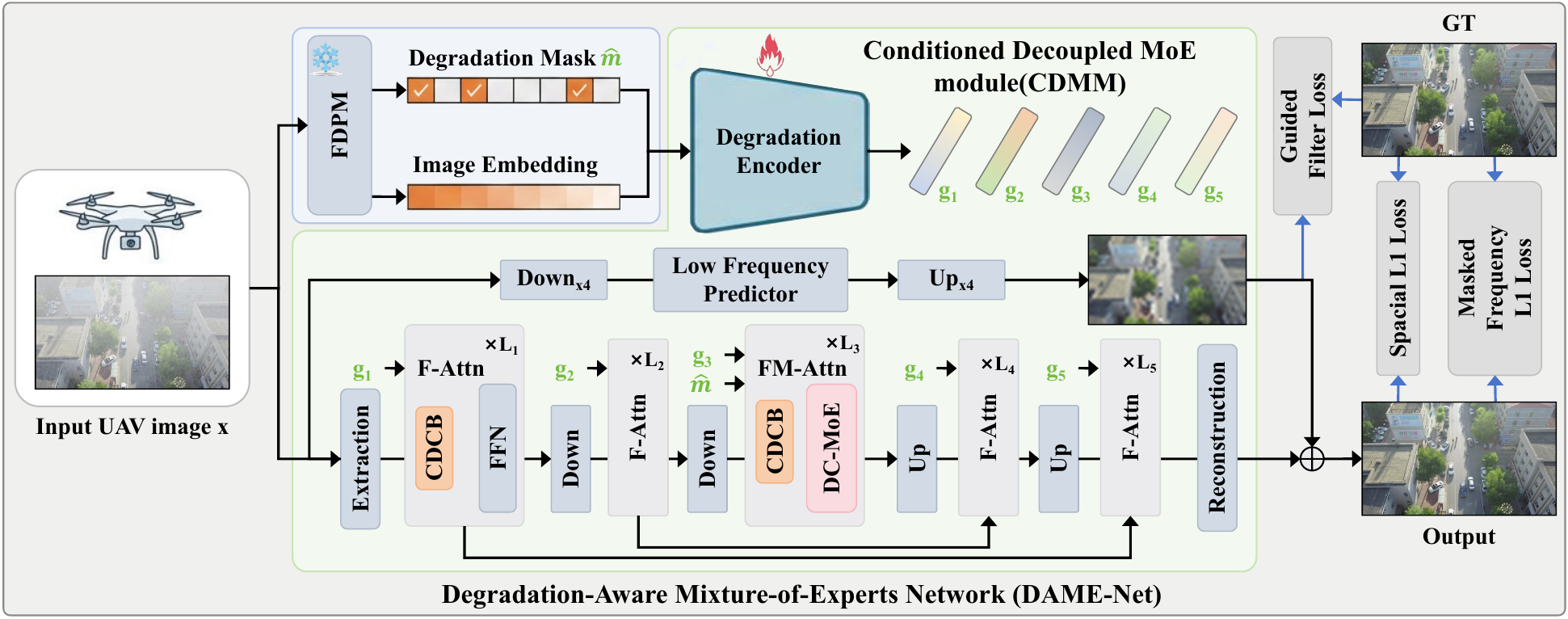}
    \caption{Overview of DAME-Net. FDPM predicts a degradation mask $\hat{m}$ and an image embedding $p$, which are transformed by the Degradation Encoder into stage-wise conditioning vectors $\{g_s\}_{s=1}^{5}$. CDMM then performs degradation-conditioned restoration, while the Low Frequency Predictor is supervised by Guided Filter Loss and the residual path is supervised by Spatial $\ell_1$ Loss and Masked Frequency $\ell_1$ Loss.}
    \label{framework}
\end{figure*}

\section{Method}
\label{sec:method}

\subsection{Problem Formulation}
We study UAV image restoration under compositional degradations, where several degradation factors may co-occur in the same observation. Let $x \in \mathbb{R}^{3\times H\times W}$ denote a degraded UAV image and $y \in \mathbb{R}^{3\times H\times W}$ its corresponding clean target. We define a set of $D$ atomic degradations $\mathcal{D}=\{d_1,\dots,d_D\}$, where $D=8$ in our setting: rain, snow, haze, low-light, over-exposure, blur, noise, and artifact. Each input is associated with a multi-hot indicator $m\in\{0,1\}^{D}$, where $m_j=1$ means that degradation $d_j$ is present.

Our key design choice is to represent each degradation configuration as a combination of atomic factors rather than as one opaque category. Restoration is therefore conditioned on an explicit per-factor description, leading to the following mapping:
\begin{equation}
\hat{y}=R(x;m,p),
\end{equation}
where $p\in\mathbb{R}^{d}$ is a semantic embedding extracted from a vision--language model (CLIP~\cite{radford2021learning}). In the full problem setting, however, the degradation composition is unknown at test time, so both the degradation description and the restoration result must be inferred from the degraded input.

\subsection{Overview}
A central difficulty in compositional restoration is that degradation perception and restoration inversion are tightly coupled: a blind model must infer the active factors and correct them with the same latent representation. We address this difficulty by explicitly decoupling degradation-aware conditioning from image reconstruction:
\begin{equation}
(\hat{m},p)=P(x), \qquad \hat{y}=R(x;\hat{m},p).
\end{equation}
Here $P$ predicts an explicit degradation mask $\hat{m}$ together with a semantic embedding $p$, and $R$ uses this conditioning instead of relying on a blind, entangled representation. Intuitively, active mask bits enable the correction pathways relevant to the detected degradations, while inactive bits suppress irrelevant processing.

$P$ is implemented by FDPM, a CLIP-based multi-label degradation detector designed for compositional degradation perception. $R$ is implemented by CDMM, a degradation-conditioned U-shaped Transformer with stage-wise conditioning, spatial--frequency hybrid processing, mask-constrained expert routing, and a low-frequency base branch for coarse illumination correction. Fig.~\ref{framework} provides an overview of the full pipeline.

\subsection{Factor-wise Degradation Perception Module (FDPM)}
FDPM infers which degradation factors are present before restoration begins by predicting at the atomic-factor level rather than treating each degradation configuration as an independent category. It outputs two signals for the restoration stage: an interpretable multi-hot mask $\hat{m}$ and a semantic embedding $p$.

Training proceeds in two stages. In Stage~I, we fine-tune the CLIP visual encoder together with a lightweight multi-label prediction head while keeping the text encoder frozen. In Stage~II, the trained FDPM is frozen and used to provide $\hat{m}$ and $p$ for restoration.

\subsubsection{Multi-Label Formulation and Aligned Multi-Task Views}
We formulate degradation perception as a multi-label prediction problem over atomic degradation factors. Specifically, we consider $K=22$ aligned training tasks $\{\tau_k\}_{k=1}^{K}$ (1 clean + 21 seen degradation configurations). Each task $\tau_k$ is represented by a multi-hot label vector $t_k\in\{0,1\}^{\hat{C}}$, where $\hat{C}=D+1$ includes an additional clean bit. During restoration, we discard the clean bit and use the remaining $D=8$ entries as the degradation mask. Although FDPM is trained only on these 22 tasks, its shared image-text embedding space also enables evaluation over all 44 valid \dataset{} configurations by instantiating configuration prompts for the combined set of 1 clean, 21 seen, and 22 unseen cases.

To reduce scene-content bias, we train FDPM on aligned multi-task views. For each sampled scene, all $K$ task variants are cropped with the same random window ($224\times224$) and processed jointly. This strategy keeps scene content fixed while varying only degradation composition, forcing the encoder to focus on corruption cues rather than semantic scene differences.

\subsubsection{CLIP-Based Multi-Label Perception}
We build FDPM on top of CLIP because its shared image-text embedding space provides a useful semantic prior for related degradation combinations. Let $E_v(\cdot)$ and $E_t(\cdot)$ denote the CLIP ViT-B/32 image and text encoders ($d=512$). Given an input image $x$, we first extract an image embedding $f_i=E_v(x)\in\mathbb{R}^{d}$ and then feed it into a lightweight multi-label head $h(\cdot)$, implemented as an MLP with LayerNorm and two linear layers (hidden width $2d$), to predict degradation logits $z=h(f_i)\in\mathbb{R}^{\hat{C}}$.

Text prompts are encoded once offline; each task $\tau_k$ uses a description such as ``This image contains rain and haze.'' At inference time, the predicted logits are converted into a hard degradation mask:
\begin{equation}
\hat{m}=\mathbf{1}\big(\sigma(z_{1:D})\ge 0.5\big).
\end{equation}
This mask provides the explicit per-factor condition used by the restoration stage.

\subsubsection{Label-Similarity-Guided Cross-Modal Alignment}
Standard contrastive alignment treats each degradation configuration as an independent class. This is suboptimal for composite degradations because it forces a mixed case to align with exactly one prompt and ignores its overlap with constituent factors. To preserve this compositional structure, we construct a label similarity matrix $S\in\mathbb{R}^{K\times K}$ from the cosine similarity between task label vectors:
\begin{equation}
S_{ij}=\frac{t_i^\top t_j}{\|t_i\|_2\|t_j\|_2}.
\end{equation}
$S$ measures the degree of factor overlap between any two configurations. For example, rain+haze should be closer to rain and haze than to noise. We therefore use $S$ to define soft cross-modal targets rather than one-hot targets.

For aligned image views and text prompts, we compute an image--text similarity matrix $A=\frac{1}{\tau}\cdot\mathrm{Norm}(F_i)\,\mathrm{Norm}(F_t)^\top$ with temperature $\tau=0.07$, and define the following soft alignment targets:
\begin{equation}
Q_{i\rightarrow t}=\mathrm{softmax}(\alpha S),\qquad Q_{t\rightarrow i}=\mathrm{softmax}(\alpha S^\top),
\end{equation}
with $\alpha=2.0$. The resulting label-similarity-guided cross-modal alignment loss is
\begin{equation}
\mathcal{L}_{\mathrm{align}}=
\frac{1}{2}\Big(
\mathrm{KL}(P_{i\rightarrow t}\,\|\,Q_{i\rightarrow t})+
\mathrm{KL}(P_{t\rightarrow i}\,\|\,Q_{t\rightarrow i})
\Big),
\end{equation}
where $P_{i\rightarrow t}=\mathrm{softmax}(A)$ and $P_{t\rightarrow i}=\mathrm{softmax}(A^\top)$. In addition to this alignment term, we apply a standard multi-label BCE loss $\mathcal{L}_{\mathrm{cls}}$ on the predicted logits. The final perception objective is
\begin{equation}
\mathcal{L}_{P}=\lambda_{\mathrm{align}}\mathcal{L}_{\mathrm{align}}+\lambda_{\mathrm{cls}}\mathcal{L}_{\mathrm{cls}}.
\label{eq:loss_perception}
\end{equation}
This design encourages FDPM to preserve compositional similarity across degradation configurations while still making accurate factor-wise predictions.

\subsection{Conditioned Decoupled MoE Module (CDMM)}
Given the degradation mask and semantic embedding from FDPM, the restoration network should emphasize the correction pathways relevant to the detected factors while suppressing irrelevant ones. CDMM is designed around this requirement. It contains four coordinated components: a degradation token encoder that converts $\hat{m}$ and $p$ into stage-wise conditioning vectors, a spatial--frequency hybrid backbone built around the Condition-Aware Dual-domain Correction Block (CDCB), a Decoupled MoE feed-forward module (DC-MoE) that first activates degradation-relevant experts according to the predicted degradation mask and then routes computation within the activated subset, and a low-frequency base branch for coarse illumination correction.

\begin{figure}[t]
    \centering
    \includegraphics[width=\linewidth]{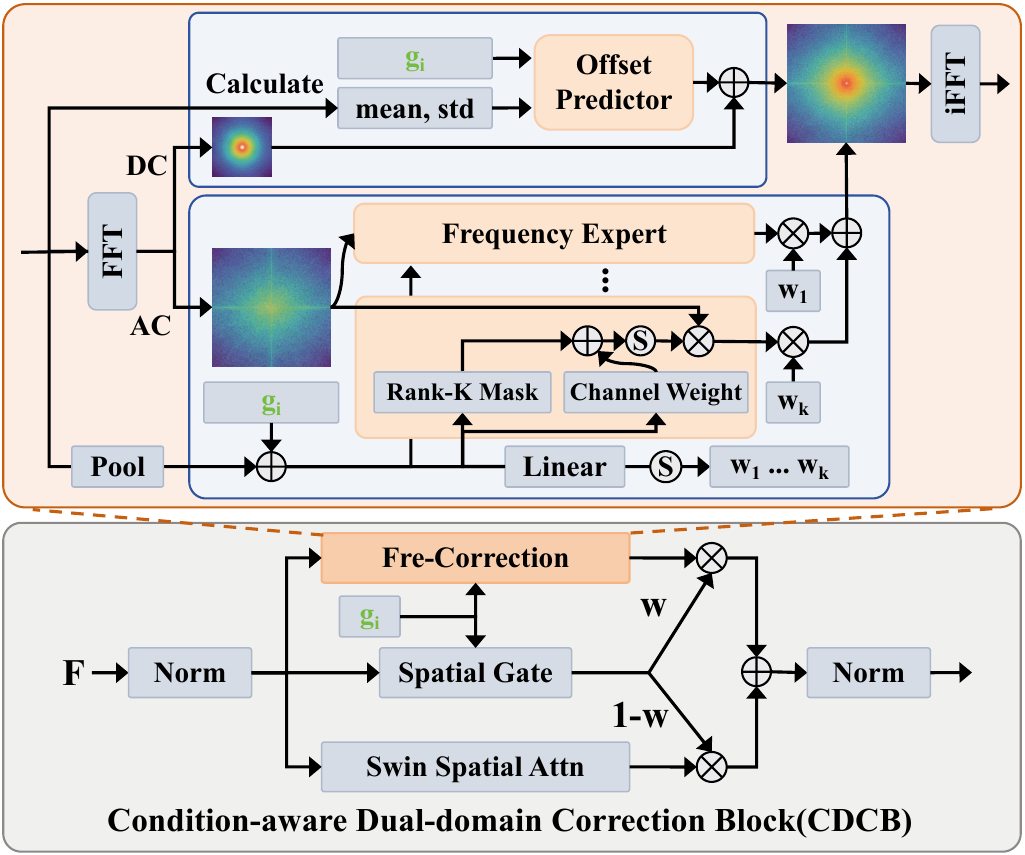}
    \caption{Architecture of CDCB. The frequency branch transforms input features via FFT, applies degradation-conditioned spectral modulation with a rank-$K$ frequency mask, and refines the DC component via content-adaptive offset prediction. The spatial branch applies Swin window attention. A learned gate $w$ balances the two branches.}
    \label{CDCB}
\end{figure}

\subsubsection{Degradation Token Encoder}
A single shared condition vector is often too coarse for compositional restoration, because different stages of the backbone may require different mixtures of global and local degradation cues. We therefore convert the FDPM outputs into stage-wise conditioning vectors $\{g_s\}_{s=1}^{S}$ for a multi-stage backbone with $S=5$ stages.

We maintain learnable degradation tokens $\{u_j\}_{j=1}^{D}$ ($u_j\in\mathbb{R}^{e}$, $e=256$), stage query tokens $\{q_s\}_{s=1}^{S}$, and two additional conditioning tokens: a semantic token $u_p=\mathrm{LN}(W_p p)$ and a global token $u_g=\mathrm{LN}(\mathrm{MLP}_g([\hat{m},p]))$. These are concatenated into the key--value set $U=[u_1;\dots;u_D;u_p;u_g]\in\mathbb{R}^{(D+2)\times e}$.

\noindent\textbf{Strict Token Masking.} A remaining issue is that absent degradations can still interfere if their tokens are allowed to participate in attention. To avoid this, tokens with $\hat{m}_j=0$ are excluded through a hard key padding mask rather than softly down-weighted. Stage-wise queries $Q=[q_1;\dots;q_S]$ then attend to $U$ via multi-head cross-attention (4 heads), followed by an FFN, yielding stage-wise conditioning vectors $g_s$ as the $s$-th row of $Z=\mathrm{Attn}(Q,U,U)\in\mathbb{R}^{S\times e}$. This produces stage-specific conditioning while reducing interference from irrelevant factors.

\subsubsection{Condition-Aware Dual-domain Correction Block (CDCB)}
Composite degradations affect both spatial structures and spectral statistics. We therefore design CDCB to process features jointly in the frequency and spatial domains, as illustrated in Fig.~\ref{CDCB}. Given input features $X\in\mathbb{R}^{C\times H\times W}$ and a stage-wise conditioning vector $g$, CDCB applies two parallel branches and fuses their outputs with a learned gate.

\begin{figure}[t]
    \centering
    \includegraphics[width=\linewidth]{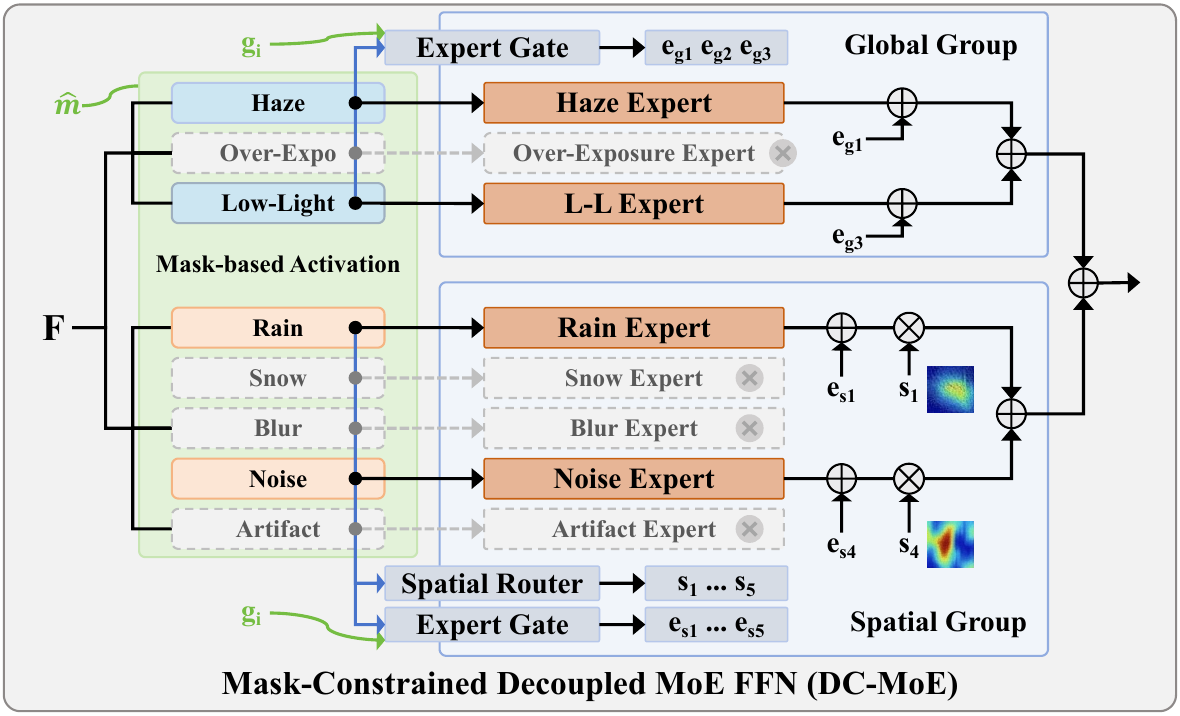}
    \caption{Architecture of the DC-MoE. The predicted degradation mask first activates degradation-relevant experts in a Global Group and a Spatial Group, while suppressing irrelevant experts. Routing weights are then assigned only over the activated expert subset. The Spatial Group further employs a pixel-wise Spatial Router to localize expert responses within the feature map.}
    \label{DCMOE}
\end{figure}

\noindent\textbf{Frequency Branch.} Degradations such as blur, noise, and compression artifacts have strong spectral signatures, so the frequency branch models them explicitly in the frequency domain. Given features $X$ and conditioning vector $g$, the branch first transforms the features with a 2D FFT and then predicts degradation-conditioned mixture weights:
\begin{equation}
\pi=\mathrm{softmax}(W_\pi(g+W_f\mathrm{GAP}(X)))\in\mathbb{R}^{M},
\end{equation}
with $M=2$ frequency experts. Each expert $m$ predicts a spectral modulation map via a low-rank factorization (rank $r=4$):
\begin{equation}
M^{(m)}=\sigma\!\left(c^{(m)}+\sum_{\ell=1}^{r} v^{(m)}_{h,\ell}\otimes v^{(m)}_{w,\ell}\right),
\end{equation}
where the outer product yields a rank-$r$ frequency mask. The modulated spectrum is inverted and experts are aggregated as
\begin{equation}
X_{\mathrm{freq}}=W_{\mathrm{out}}\sum_{m}\pi_m\cdot\mathcal{F}^{-1}_{2D}(\tilde{X}\odot M^{(m)}).
\end{equation}

\noindent\textbf{Content-Adaptive DC Correction.} Spectral modulation alone is insufficient for degradations dominated by global illumination shifts, such as haze, low-light, and over-exposure. We therefore refine the DC component explicitly. Let $\mu$ and $\sigma$ denote the per-channel mean and standard deviation of $X$. We update the zero-frequency entry as:
\begin{equation}
M^{(m)}_{:,0,0} \leftarrow 1+b_{dc}+\eta\cdot\tanh(\mathrm{MLP}_{dc}([g,\mu,\sigma])),
\end{equation}
where $\eta=0.1$ bounds the correction magnitude. This allows the frequency branch to address both local high-frequency artifacts and global illumination changes without introducing a separate post-processing module.

\noindent\textbf{Spatial Branch and Gating.} The spatial branch applies window-based Swin attention to capture local structural correlations. A learned scalar gate $w\in[0,1]$ then balances the frequency and spatial branches:
\begin{equation}
X_{\mathrm{out}} = w \cdot X_{\mathrm{freq}} + (1-w) \cdot X_{\mathrm{spatial}},
\end{equation}
where $X_{\mathrm{spatial}}$ is the output of the Swin attention branch. This gate allows CDCB to shift emphasis toward frequency processing for spectrally prominent degradations such as blur and noise, and toward spatial processing for structurally localized ones such as rain streaks.

\subsubsection{Decoupled MoE Feed-Forward (DC-MoE)}
To reduce interference among heterogeneous corrections, we decouple the feed-forward experts into two groups: $E_g=3$ global experts for scene-level degradations (haze, low-light, over-exposure) and $E_s=5$ spatial experts for localized degradations (rain, snow, blur, noise, artifact). Given a feature map $X$, the predicted degradation mask first selects the candidate experts in the global and spatial groups. Two independent gates then assign group-specific routing weights over the activated subset, followed by renormalization:
\begin{equation}
\hat{\pi}^{g}=\mathrm{Renorm}(\pi^{g}\odot \hat{m}^{g}), \qquad
\hat{\pi}^{s}=\mathrm{Renorm}(\pi^{s}\odot \hat{m}^{s}),
\end{equation}
where $\hat{m}^{g}$ selects the \{haze, low-light, over-exposure\} bits and $\hat{m}^{s}$ selects the \{rain, snow, blur, noise, artifact\} bits. For spatial experts, we further predict a spatial routing map $\mathcal{R}_j\in[0,1]^{H\times W}$ to localize expert responses. The resulting decoupled MoE FFN is
\begin{equation}
\mathrm{FFN}_{\mathrm{MoE}}(X)=B(X)+\sum_{i}\hat{\pi}^{g}_{i}E^{g}_{i}(X)+\sum_{j}\hat{\pi}^{s}_{j}\,\mathcal{R}_j\odot E^{s}_{j}(X),
\end{equation}
where $B(\cdot)$ is a base branch that provides non-zero capacity even when no expert is activated. This design lets global and local degradations use different expert pools while preventing absent degradation types from consuming routing capacity.

\subsubsection{Base-Residual Dual-Branch Reconstruction}
A final issue is that coarse illumination correction and fine-detail recovery place different demands on the decoder. We therefore adopt a dual-branch reconstruction design:
\begin{equation}
\hat{y}=\hat{y}_{\mathrm{base}}+\hat{y}_{\mathrm{res}},
\end{equation}
The base branch predicts a low-frequency layer $\hat{y}_{\mathrm{base}}$ by downsampling the input $4\times$ (bilinear), processing it with a lightweight CNN at low resolution, and upsampling back. Since coarse structural and illumination corrections are predominantly low-frequency, this branch handles them efficiently. The Transformer backbone then predicts the high-frequency residual $\hat{y}_{\mathrm{res}}$ at full resolution. This separation stabilizes training and reduces the tendency of the residual path to absorb coarse illumination errors.

\subsection{Training Objectives}
We train the framework in two stages. The perception model $P$ is trained first using Eq.~\eqref{eq:loss_perception}; once converged, it is frozen for all subsequent restoration training.

The restoration objective is
\begin{equation}
\mathcal{L}_{R}= \|\hat{y}-y\|_{1}+\lambda_f \mathcal{L}_{\mathrm{freq}}+\lambda_p \mathcal{L}_{\mathrm{base}},
\end{equation}
with $(\lambda_f,\lambda_p)=(0.1,0.1)$. Here $\mathcal{L}_{\mathrm{freq}}$ is a masked FFT-magnitude $\ell_1$ loss: a square low-frequency center region (ratio 0.2 of the shorter side) is removed before computing the spectral loss so that supervision focuses on mid-to-high frequency content, where blur, noise, and compression artifacts are most prominent. $\mathcal{L}_{\mathrm{base}}$ supervises the base branch against a guided-filter smoothed target $y_{\mathrm{base}}=\mathrm{GuidedFilter}(y,y;r=15,\epsilon=10^{-3})$.

We additionally introduce mask-overload augmentation. With probability 0.05, samples containing only rain or snow (with no haze or low-light flag active) are assigned one randomly activated global degradation bit. This perturbation reduces over-reliance on a perfectly accurate mask by forcing the routing mechanism to suppress irrelevant global experts when the image content does not support the extra activation.

\begin{table*}[t]
\centering
\caption{Quantitative comparison on \dataset{} using averaged results only (PSNR/SSIM). Results are reported for in-distribution (seen) and zero-shot (unseen) degradation settings. Best and second-best results are marked in \textbf{bold} and \underline{underlined}, respectively.}
\label{tab:comparison_avg_merged}
\resizebox{\textwidth}{!}{%
\begin{threeparttable}
\begin{tabular}{lcccccccccccc}
\toprule
\multirow{2}{*}{Setting}
  & \multicolumn{2}{c}{\textbf{Ours}}
  & \multicolumn{2}{c}{PromptIR~\cite{potlapalli2023promptir}}
  & \multicolumn{2}{c}{DehazeFormer~\cite{song2023vision}}
  & \multicolumn{2}{c}{AirNet~\cite{li2022all}}
  & \multicolumn{2}{c}{Restormer~\cite{zamir2022restormer}}
  & \multicolumn{2}{c}{AdaIR~\cite{cui2025adair}} \\
\cmidrule(lr){2-3}\cmidrule(lr){4-5}\cmidrule(lr){6-7}\cmidrule(lr){8-9}\cmidrule(lr){10-11}\cmidrule(lr){12-13}
  & PSNR & SSIM & PSNR & SSIM & PSNR & SSIM & PSNR & SSIM & PSNR & SSIM & PSNR & SSIM \\
\midrule
\multicolumn{13}{l}{\textit{Seen Settings}} \\
\midrule
Avg. Seen Single  & \textbf{29.52} & \textbf{0.9091} & \underline{29.36} & \underline{0.9066} & 28.93 & 0.9002 & 27.40 & 0.8347 & 29.16 & 0.9041 & 28.40 & 0.8931 \\
Avg. Seen Double  & \textbf{26.88} & \textbf{0.8389} & \underline{26.61} & \underline{0.8328} & 26.22 & 0.8207 & 25.07 & 0.7454 & 26.43 & 0.8283 & 25.79 & 0.8102 \\
Avg. Seen Triple  & \textbf{25.73} & \textbf{0.8105} & \underline{25.39} & \underline{0.7988} & 24.95 & 0.7804 & 24.36 & 0.7048 & 25.25 & 0.7927 & 24.65 & 0.7681 \\
Overall Seen      & \textbf{27.67} & \textbf{0.8602} & \underline{27.43} & \underline{0.8544} & 27.01 & 0.8433 & 25.82 & 0.7717 & 27.25 & 0.8504 & 26.57 & 0.8338 \\
\midrule
\multicolumn{13}{l}{\textit{Zero-Shot Settings}} \\
\midrule
Avg. Unseen Double & \textbf{20.46} & \textbf{0.7219} & \underline{18.60} & \underline{0.6681} & 18.54 & 0.6534 & 17.82 & 0.5949 & 18.48 & 0.6635 & 18.19 & 0.6460 \\
Avg. Unseen Triple & \textbf{18.16} & \textbf{0.6524} & 17.73 & 0.6502 & 17.84 & 0.6336 & 17.70 & 0.5946 & \underline{17.91} & \underline{0.6514} & 17.90 & 0.6343 \\
Avg. Unseen Quad   & \textbf{16.98} & \textbf{0.5211} & 13.77 & \underline{0.4671} & 13.45 & 0.4412 & 13.49 & 0.4254 & \underline{13.78} & 0.4590 & 13.66 & 0.4525 \\
Overall Unseen     & \textbf{18.62} & \textbf{0.6271} & \underline{16.46} & \underline{0.5826} & 16.33 & 0.5630 & 16.03 & 0.5255 & 16.45 & 0.5776 & 16.28 & 0.5647 \\
\bottomrule
\end{tabular}
\end{threeparttable}
}
\end{table*}

\begin{table}[t]
\centering
\caption{Downstream object detection on all 43 degraded UAV settings. ``Degraded'' denotes detection on degraded inputs, ``GT'' the upper bound on clean images, and $\Delta$mAP50 the improvement over Degraded. Best and second-best results are marked in \textbf{bold} and \underline{underlined}, respectively.}
\label{tab:downstream_detection}
\resizebox{\columnwidth}{!}{%
\begin{tabular}{lccccc}
\toprule
Method & Precision$\uparrow$ & Recall$\uparrow$ & mAP50$\uparrow$ & mAP50-95$\uparrow$ & $\Delta$mAP50$\uparrow$ \\
\midrule
GT     & \textbf{0.5851} & \textbf{0.5486} & \textbf{0.5419} & \textbf{0.3211} & -- \\
Degraded & 0.1530 & 0.1046 & 0.0971 & 0.0542 & -- \\
\midrule
AdaIR         & 0.3189 & 0.2343 & 0.2216 & 0.1259 & 0.1245 \\
DehazeFormer  & 0.3269 & 0.2392 & 0.2289 & 0.1307 & 0.1318 \\
Restormer     & 0.3434 & 0.2505 & 0.2420 & 0.1380 & 0.1449 \\
PromptIR      & 0.3495 & 0.2547 & 0.2469 & 0.1412 & \underline{0.1497} \\
Ours & \underline{0.3530} & \underline{0.2603} & \underline{0.2518} & \underline{0.1444} & \textbf{0.1547} \\
\bottomrule
\end{tabular}
}
\end{table}

\section{Experiments}
\label{sec:exp}

\subsection{Experimental Setup}
\label{sec:exp-setup}

\noindent\textbf{Dataset and splits.}
All experiments are conducted on \dataset{}, which covers 8 atomic degradation types---rain, snow, haze, low-light, over-exposure, blur, noise, and artifact---and composes up to 4 factors within one image. This yields 43 valid configurations, each annotated with a multi-hot label for explicit compositional supervision. We train on 21 configurations and hold out the remaining 22 for zero-shot evaluation under a standardized seen/unseen split. Much of the unseen split centers on low-light+blur and low-light+artifact, plus their higher-order extensions, to test generalization to jointly unseen combinations. Detailed construction, synthesis rules, and benchmark comparison are deferred to the supplementary material.

\noindent\textbf{Evaluation metrics.}
We report PSNR (dB) and SSIM on the luminance channel (Y channel in YCbCr space). To highlight performance trends under different restoration difficulties, the main paper summarizes the results using group averages organized by degradation complexity (single, double, triple, and quad) and by evaluation protocol (seen versus zero-shot unseen), while the full task-wise tables are reported in the supplementary material.

\noindent\textbf{Baseline methods.}
We compare against five representative unified restoration methods: AirNet~\cite{li2022all}, DehazeFormer~\cite{song2023vision}, Restormer~\cite{zamir2022restormer}, PromptIR~\cite{potlapalli2023promptir}, and AdaIR~\cite{cui2025adair}. All baselines are retrained on the same \dataset{} training split under identical protocols for fair comparison.

\noindent\textbf{Implementation details.}
FDPM is initialized from CLIP ViT-B/32. In Stage~I, we fine-tune the visual encoder and multi-label head while freezing the text encoder; in Stage~II, FDPM is frozen and provides $\hat{m}$ and $p$ to the restoration network. We use $K=22$ aligned multi-task views and $(\lambda_{\mathrm{align}},\lambda_{\mathrm{cls}})=(0.1,0.9)$. CDMM adopts a 5-stage U-shaped architecture with channel widths [24, 48, 96, 48, 24], $M=2$ frequency experts with rank $r=4$, and $E_g=3$ global/$E_s=5$ spatial MoE experts. The restoration model is trained with AdamW (lr=$2\times10^{-4}$, weight decay~0.02), batch size~8, crop size $256\times256$, and 100 epochs, with $(\lambda_f,\lambda_p)=(0.1,0.1)$ and mask-overload probability 0.05. All experiments use 1$\times$ NVIDIA A100 GPU. Additional detector evaluation and complexity analysis are deferred to the supplementary material.

\subsection{Comparison with State-of-the-Art Methods}
\label{sec:exp-comparison}

\subsubsection{Quantitative Results}
Table~\ref{tab:comparison_avg_merged} summarizes the averaged PSNR and SSIM results on \dataset{} under both seen and zero-shot unseen settings. Our method ranks first across all reported group averages, showing consistent advantages over strong unified restoration baselines. The corresponding task-wise seen and zero-shot results are provided in the supplementary material.

On the seen settings, the improvement is moderate but stable. Compared with PromptIR, our method improves the overall seen average from 27.43 dB / 0.8544 to 27.67 dB / 0.8602, and the gains increase with degradation complexity: from 29.52 dB / 0.9091 on seen single degradations, to 26.88 dB / 0.8389 on seen double degradations, and 25.73 dB / 0.8105 on seen triple degradations. This trend suggests that even in-distribution, explicit per-factor conditioning becomes more beneficial as multiple degradation factors must be handled jointly.

The advantage becomes more pronounced on the zero-shot unseen settings. Our method achieves 20.46 dB / 0.7219 on unseen double degradations, 18.16 dB / 0.6524 on unseen triple degradations, and 16.98 dB / 0.5211 on unseen quad degradations, with an overall unseen average of 18.62 dB / 0.6271. The larger margin on unseen quad degradations indicates that the proposed DAME-Net design generalizes better when several degradation factors appear together in combinations never observed during training.

\noindent\textbf{Limitations.}
The averaged results also reveal that zero-shot compositional restoration remains challenging for all methods, especially as the number of active degradation factors increases. Although our method achieves the best performance across all reported averages, the absolute restoration quality still drops substantially from seen to unseen settings and from lower-order to higher-order composites. This gap suggests that heavily coupled degradations and severe visibility loss remain difficult for the current explicit degradation representation. Improving factor separability in FDPM and strengthening restoration robustness under more complex mixtures are therefore important directions for future work.

\subsubsection{Qualitative Comparison}
\begin{figure*}[t]
    \centering
    \includegraphics[width=0.9\linewidth]{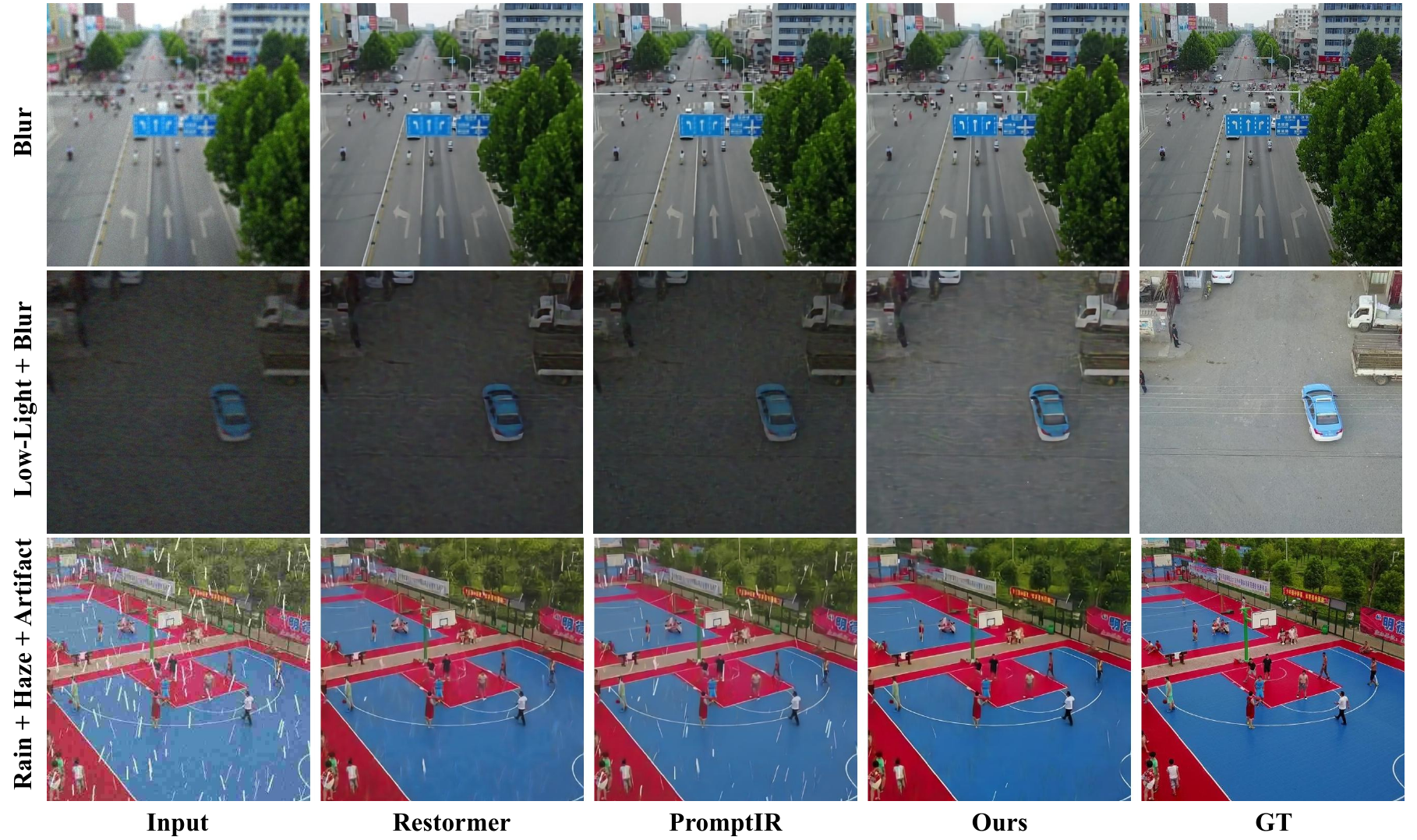}
    \caption{Qualitative comparison on \dataset{} under single and composite degradations. From left to right: Input, Restormer, PromptIR, Ours, and Ground Truth (GT). Our method better preserves scene structure and color consistency while leaving fewer visible residual degradations.}
    \label{fig:qualitative}
\end{figure*}

Fig.~\ref{fig:qualitative} provides representative visual comparisons under single and composite degradations, complementing the averaged quantitative trends in Table~\ref{tab:comparison_avg_merged} and the task-wise results reported in the supplementary material. On the seen single-degradation case of blur, our method preserves object boundaries more clearly and leaves fewer residual artifacts than the competing methods. The advantage becomes more pronounced on unseen configurations such as low-light+blur and rain+haze+artifact, where competing methods often leave residual blur, color distortion, or under-enhanced dark regions, whereas our results recover cleaner local structure and more balanced overall appearance. These cases require joint correction of global illumination distortion and local detail corruption, which is better supported by our explicit per-factor conditioning and spatial--frequency restoration design. This visual evidence is consistent with the larger zero-shot gains summarized in Table~\ref{tab:comparison_avg_merged}.

\subsection{Downstream Object Detection}
We further evaluate whether the restoration gains on \dataset{} translate into downstream UAV perception. A frozen pre-trained YOLOv8n detector is applied to degraded inputs, clean references, and restored outputs from all methods across all 43 configurations. We report Precision, Recall, mAP50, and mAP50-95, and define $\Delta$mAP50 as the gain over degraded inputs.

Composite degradations severely harm detection, with mAP50 dropping from 0.5419 on clean images to 0.0971 on degraded inputs. All restoration methods improve downstream detection, and our method performs best overall, reaching 0.2518 mAP50 and a 0.1547 gain over degraded inputs. This result is consistent with the restoration improvements reported above and further supports the practical value of compositional UAV image restoration for downstream UAV perception.

\subsection{Ablation Study}
\label{sec:exp-ablation}

We ablate each major component by removing or replacing it from the full model and evaluating on the full \dataset{} test set. We report mean PSNR/SSIM over all 43 tasks and, separately, over the quad-degradation subset where performance gaps are most amplified. The ablations are organized into three groups corresponding to our pipeline design: degradation representation, restoration architecture, and training strategy.

\subsubsection{Degradation Representation Modules}

Table~\ref{tab:ablation-perception} ablates the components of FDPM and the degradation token encoder.

Removing the CLIP image embedding $p$ (``w/o Semantic Embedding'') produces the largest overall drop: 1.46 dB on average and 3.31 dB on quad tasks. This indicates that the semantic embedding provides information beyond the binary mask, especially when four factors must be handled jointly. This interpretation is further supported by the oracle-mask check below: replacing the predicted binary mask with the ground-truth mask changes the result only marginally. Removing the global token $u_g$ costs 1.08 dB overall and 2.57 dB on quad.

Removing strict token masking (``w/o Strict Token Masking'') reduces overall performance by 0.56 dB, indicating that excluding inactive degradation tokens is beneficial in this setting. Substituting a soft mask (``Soft Mask'') recovers most of this loss, which suggests that the masking strategy mainly affects robustness across different task configurations.

\begin{table}[t]
\centering
\caption{Ablation of degradation representation components. ``GT Mask'' replaces the predicted binary mask with the ground-truth mask while keeping the CLIP-derived semantic embedding unchanged. ``All'' and ``Quad'' denote averages over all 43 tasks and the quad-degradation subset, respectively. Best and second-best results are marked in \textbf{bold} and \underline{underlined}, respectively.}
\label{tab:ablation-perception}
\begin{tabular}{lcccc}
\toprule
\multirow{2}{*}{Variant} & \multicolumn{2}{c}{All} & \multicolumn{2}{c}{Quad} \\
\cmidrule(lr){2-3}\cmidrule(lr){4-5}
 & PSNR & SSIM & PSNR & SSIM \\
\midrule
Full model (Ours)              & \textbf{23.04} & \underline{0.7410} & 16.98 & 0.5211 \\
GT Mask (oracle mask)          & \textbf{23.04} & \textbf{0.7415} & 16.97 & \underline{0.5226} \\
\midrule
w/o Semantic Embedding         & 21.58 & 0.7055 & 13.67 & 0.4356 \\
w/o Global Token               & 21.96 & 0.7201 & 14.41 & 0.4707 \\
w/o Strict Token Masking       & 22.48 & 0.7320 & 15.96 & 0.5060 \\
Soft Mask (replace hard mask)  & \underline{22.89} & 0.7390 & \underline{17.05} & 0.5221 \\
w/o Semantic Token             & 22.80 & 0.7400 & 16.60 & \textbf{0.5229} \\
w/o Stage-wise Embed           & 22.74 & 0.7321 & \textbf{17.27} & 0.5144 \\
\bottomrule
\end{tabular}
\end{table}

Removing the semantic token $u_p$ (``w/o Semantic Token'') costs 0.24 dB, while replacing stage-wise conditioning with a single shared embedding (``w/o Stage-wise Embed'') costs 0.30 dB. These smaller but consistent drops suggest that hierarchical conditioning improves how degradation cues are distributed across the backbone.

We further conduct an oracle-mask check by replacing only the predicted binary degradation mask with the ground-truth mask while keeping the CLIP-derived semantic embedding unchanged. This changes the full 43-task result only marginally, from 23.04 dB / 0.7410 SSIM to 23.04 dB / 0.7415 SSIM. The same near-zero gap also holds on the seen, zero-shot, and quad subsets, where the oracle-mask variant reaches 27.67 / 0.8604, 18.62 / 0.6279, and 16.97 / 0.5226, respectively. This indicates that binary mask prediction is not the dominant bottleneck in the current system; the remaining errors are more likely due to restoration difficulty and finer-grained conditioning quality beyond the hard mask itself.

\subsubsection{Restoration Architecture}

Table~\ref{tab:ablation-arch} ablates the architectural components of CDMM.

Removing the frequency branch of CDCB entirely (``w/o Freq Branch'') costs 0.76 dB overall. Removing the spatial--frequency gate (``w/o Freq-Spa Gate'') leads to a larger drop: 0.95 dB overall and 1.76 dB on quad tasks. This result suggests that balancing spatial and spectral processing is more important than using the frequency branch alone in a fixed way.

Replacing the decoupled MoE with a standard fully-shared MoE (``w/o DC-MoE'') costs 0.70 dB overall and 1.50 dB on quad tasks, indicating that expert separation by degradation type is useful in the compositional setting. Removing the decoupling gate (``w/o Decouple Gate'') causes a further 0.58 dB drop, while removing the spatial routing map (``w/o Spatial Router'') incurs a smaller but consistent 0.28 dB penalty.

Removing DC correction reduces PSNR by 0.40 dB overall, with the effect concentrated on degradations dominated by global illumination changes. Removing the base-residual dual branch costs 0.47 dB, suggesting that separating coarse low-frequency correction from high-frequency residual recovery remains beneficial.

\begin{table}[t]
\centering
\caption{Ablation of restoration architecture components. ``All'' and ``Quad'' denote averages over all 43 tasks and the quad-degradation subset, respectively. Best and second-best results are marked in \textbf{bold} and \underline{underlined}, respectively.}
\label{tab:ablation-arch}
\begin{tabular}{lcccc}
\toprule
\multirow{2}{*}{Variant} & \multicolumn{2}{c}{All} & \multicolumn{2}{c}{Quad} \\
\cmidrule(lr){2-3}\cmidrule(lr){4-5}
 & PSNR & SSIM & PSNR & SSIM \\
\midrule
Full model (Ours)        & \textbf{23.04} & \textbf{0.7410} & \underline{16.98} & \underline{0.5211} \\
\midrule
w/o Freq-Spa Gate        & 22.09 & 0.7203 & 15.22 & 0.4765 \\
w/o DC-MoE               & 22.34 & 0.7222 & 15.48 & 0.4804 \\
w/o Decouple Gate        & 22.46 & 0.7284 & 15.55 & 0.4836 \\
w/o Freq Branch          & 22.28 & 0.7278 & 16.12 & 0.5108 \\
w/o Dual Branch          & 22.56 & 0.7327 & 16.52 & 0.5160 \\
w/o DC Correction        & 22.63 & 0.7328 & \textbf{17.20} & 0.5158 \\
w/o Spatial Router       & \underline{22.76} & \underline{0.7377} & 16.80 & \textbf{0.5223} \\
\bottomrule
\end{tabular}
\end{table}

\subsubsection{Training Strategy}

Table~\ref{tab:ablation-training} ablates the training objectives and mask-overload augmentation.

Dropping $\mathcal{L}_{\mathrm{freq}}$ (``w/o Freq Loss'') reduces overall PSNR by 0.72 dB and quad-task PSNR by 2.49 dB. This confirms that explicit supervision on mid-to-high frequency bands is particularly important when several detail-damaging degradations co-occur.

Removing mask-overload augmentation (``w/o Mask Overload'') costs 0.46 dB overall and 0.44 dB on quad. This indicates that explicitly training the model to cope with imperfect degradation cues improves robustness beyond the augmented samples themselves.

Removing $\mathcal{L}_{\mathrm{base}}$ (``w/o Guided Filter Loss'') reduces PSNR by 0.13 dB while increasing SSIM by 0.0043. This reveals a mild trade-off: the guided-filter target favors low-frequency structure, which benefits PSNR, whereas removing it slightly improves SSIM. Given the PSNR gain, we retain this loss in the full model.

Overall, the ablations show that the gain comes from the full pipeline rather than any single module. The largest drops arise from degradation representation, especially the semantic embedding and the global token, while the restoration architecture and training objectives provide further gains. This pattern supports the design choice of combining explicit degradation perception with degradation-conditioned restoration.

\section{Conclusion}
We address UAV image restoration under compositional degradations with DAME-Net, in which FDPM predicts an interpretable multi-hot degradation mask and CDMM restores the image through stage-wise conditioning, spatial--frequency processing, and mask-constrained expert routing. We also introduce \dataset{}, a benchmark with 43 degradation configurations and standardized seen/unseen splits. Experiments on \dataset{} show consistent improvements over representative unified restoration baselines, especially on unseen and higher-order degradations, and downstream object detection experiments further show benefits for downstream UAV perception. However, restoration quality still drops markedly from seen to unseen settings and as more degradation factors are combined, indicating that heavily coupled degradations remain challenging. Extending the framework to more realistic compositional degradations while improving degradation-perception robustness is an important direction for future work. Additional task-wise tables, detector evaluation, and complexity analysis are provided in the supplementary material.

\begin{table}[t]
\centering
\caption{Ablation of training strategies. ``All'' and ``Quad'' denote averages over all 43 tasks and the quad-degradation subset, respectively. Best and second-best results are marked in \textbf{bold} and \underline{underlined}, respectively.}
\label{tab:ablation-training}
\begin{tabular}{lcccc}
\toprule
\multirow{2}{*}{Variant} & \multicolumn{2}{c}{All} & \multicolumn{2}{c}{Quad} \\
\cmidrule(lr){2-3}\cmidrule(lr){4-5}
 & PSNR & SSIM & PSNR & SSIM \\
\midrule
Full model (Ours)          & \textbf{23.04} & \underline{0.7410} & \textbf{16.98} & \underline{0.5211} \\
\midrule
w/o Freq Loss              & 22.32 & 0.7304 & 14.49 & 0.4790 \\
w/o Mask Overload          & 22.58 & 0.7287 & 16.54 & 0.4910 \\
w/o Guided Filter Loss     & \underline{22.91} & \textbf{0.7452} & \underline{16.88} & \textbf{0.5337} \\
\bottomrule
\end{tabular}
\end{table}

\bibliographystyle{IEEEtran}
\bibliography{ref}  
\end{document}